\documentclass[10pt,conference]{IEEEtran}
\IEEEoverridecommandlockouts
\usepackage[hyphens]{url}
\usepackage{amsmath,amssymb,amsfonts}
\usepackage{algorithmic}
\usepackage{graphicx}
\usepackage{textcomp}
\usepackage{xcolor}
\usepackage{hyperref}

\usepackage{array}
\usepackage{listings}
\usepackage{multirow}
\usepackage{mdframed}
\usepackage{subcaption}
\usepackage{stfloats}
\usepackage{placeins}
\usepackage[table]{xcolor}
\usepackage{colortbl}
\usepackage{listings}
\usepackage{booktabs}
\usepackage{xcolor}
\usepackage{tabularx}
\usepackage{siunitx}
\usepackage[T1]{fontenc}

\usepackage{tcolorbox}
\usepackage{listings}
\usepackage{fvextra}
\usepackage[inkscapelatex=false]{svg}
\usepackage{float} 
\usepackage{subcaption}
\usepackage{minted}

\usepackage[table]{xcolor}

\definecolor{codegreen}{rgb}{0,0.6,0}
\definecolor{codegray}{rgb}{0.5,0.5,0.5}
\definecolor{codepurple}{rgb}{0.58,0,0.82}
\definecolor{backcolour}{rgb}{0.95,0.95,0.92}
\definecolor{bestcell}{RGB}{173, 216, 230}

\definecolor{diffold}{RGB}{254,226,226}   
\definecolor{diffoldt}{RGB}{127,29,29}    
\definecolor{diffnew}{RGB}{209,250,229}   
\definecolor{diffnewt}{RGB}{6,78,59}      
\definecolor{hlred}{RGB}{254,202,202}     
\definecolor{hlgreen}{RGB}{167,243,208}   
\definecolor{hlamber}{RGB}{254,243,199}   
\definecolor{codecomment}{RGB}{107,114,128}
\definecolor{codebg}{RGB}{249,250,251}

\lstdefinestyle{cstyle}{
  language=C,
  basicstyle=\ttfamily\scriptsize,
  commentstyle=\color{codecomment}\itshape,
  keywordstyle=\bfseries,
  backgroundcolor=\color{codebg},
  frame=none,
  breaklines=true,
  breakatwhitespace=true,
  columns=fullflexible,
  keepspaces=true,
  escapeinside={(*@}{@*)},
  xleftmargin=4pt,
  xrightmargin=4pt,
  aboveskip=2pt,
  belowskip=2pt,
}

\lstdefinestyle{ruststyle}{
  language=C,  
  basicstyle=\ttfamily\scriptsize,
  commentstyle=\color{codecomment}\itshape,
  keywordstyle=\bfseries,
  morekeywords={let, mut, fn, loop, break, println, if, for, in, use, std,
                io, Read, f64, i64, Ok, Err, match, return},
  backgroundcolor=\color{codebg},
  frame=none,
  breaklines=true,
  breakatwhitespace=true,
  columns=fullflexible,
  keepspaces=true,
  escapeinside={(*@}{@*)},
  xleftmargin=4pt,
  xrightmargin=4pt,
  aboveskip=2pt,
  belowskip=2pt,
}

\NewEnviron{opus-box}[1][]{
 \vspace{0.1cm}
    \begin{mdframed}[backgroundcolor=gray!25]
        \BODY
    \end{mdframed}
    \vspace{0.1cm}
}

\NewEnviron{prompt-box}[1][]{
 \vspace{0.1cm}
    \begin{mdframed}[backgroundcolor=yellow!25]
        \BODY
    \end{mdframed}
    \vspace{0.1cm}
}

\definecolor{bestgreen}{RGB}{198,239,206}

\def\BibTeX{{\rm B\kern-.05em{\sc i\kern-.025em b}\kern-.08em
    T\kern-.1667em\lower.7ex\hbox{E}\kern-.125emX}}
\begin{document}

\title{Beyond Source Code: Evaluating Natural Language Specifications for LLM-Based C-to-Rust Translation}


\author{
\IEEEauthorblockN{Biruk Tadesse}
\IEEEauthorblockA{
North Carolina State University\\
Raleigh, North Carolina, USA\\
btadess@ncsu.edu
}
\and
\IEEEauthorblockN{Lawrence Arkoh}
\IEEEauthorblockA{
North Carolina State University\\
Raleigh, North Carolina, USA\\
larkoh@ncsu.edu
}
\and[\hfill\mbox{}\par\mbox{}\hfill]
\IEEEauthorblockN{Marcelo d'Amorim}
\IEEEauthorblockA{
North Carolina State University\\
Raleigh, North Carolina, USA\\
mdamori@ncsu.edu
}
\and
\IEEEauthorblockN{Wesley K. G. Assun\c{c}\~ao}
\IEEEauthorblockA{
North Carolina State University\\
Raleigh, North Carolina, USA\\
wguezas@ncsu.edu
}
}



\maketitle
\begin{abstract}
Large language models (LLMs) have shown promising results for C-to-Rust translation, yet existing approaches primarily rely on source code as input, leaving natural-language specifications of program behavior underexplored as an alternative or complementary representation. We hypothesize that direct code translation encourages LLMs to preserve C-specific implementation choices, producing Rust that may be functionally correct but unnecessarily verbose, reliant on unsafe constructs, and less idiomatic. We investigate this hypothesis by comparing three translation settings: code-only, specification-only, and code-and-specification. Using 300 C programs from the IBM CodeNet dataset, we evaluate GPT-5.2 and GPT-4o-mini on functional correctness, maintainability, complexity, idiomaticity, unsafe construct usage, and cost. Code-only translation achieves the highest test pass rate (98.6\%) but often mirrors C's implementation patterns. In contrast, specification-only translation produces the most idiomatic and compact Rust, reducing unsafe blocks by approximately 6$\times$ and Clippy style warnings by 2.4$\times$. However, using specifications without validation yields test pass rates as low as 86\% with our best performing model. Adding an iterative test-based validation and refinement step for specifications raises this to a 95.3\% test pass rate. Overall, specification-based translation consistently improves code quality relative to code-only translation, while code-and-specification offers the best balance between correctness (98.3\% test pass rate) and maintainability. Our study systematically compares these translation settings, highlighting the trade-offs among specification quality, translation correctness, code quality, and computational cost.
\end{abstract}

\begin{IEEEkeywords}
Software Modernization, Code Translation, Software Maintainability, Large Language Models
\end{IEEEkeywords}

\section{Introduction}

Software written in C underpins critical systems such as operating systems, embedded firmware, and network infrastructure~\cite{vandervelden2024overview}. Although C offers high performance and fine-grained control over memory, this flexibility comes at the cost of memory safety. Buffer overflows, use-after-free errors, and null-pointer dereferences remain a major source of software vulnerabilities~\cite{10102611}. Microsoft and Google report that approximately 70\% of their critical security bugs originate from memory-unsafe languages such as C~\cite{msrc2019saferlanguage,vanderstoep2019queue}. Consequently, there is growing industrial interest in replacing legacy C with memory-safe alternatives.

Rust has emerged as one of the most promising candidates for this migration. It provides performance comparable to C while enforcing memory safety at compile time through its ownership and borrowing system~\cite{10.1145/2692956.2663188}. This has motivated large-scale migration initiatives in both industry and government~\cite{darpa2024tractor,cisa2023memorysafe,Thomas2023}. Unfortunately, manually rewriting mature C systems in Rust is prohibitively expensive and error-prone. Existing transpilers, such as C2Rust~\cite{c2rust}, automate much of the syntactic conversion but largely preserve the original implementation, often producing Rust that resembles ``C written in Rust syntax'', with excessive \texttt{unsafe} blocks, non-idiomatic constructs, and limited maintainability~\cite{emre2021translating,tadesse2026}. As a result, these tools only partially realize the benefits that motivate migration to Rust in the first place.

Large Language Models (LLMs) have recently emerged as a promising alternative. Rather than performing mechanical translation, LLMs can synthesize Rust that is more natural, idiomatic, and easier to maintain~\cite{C2SaferRust2025, 10.1145/3786180.3788317,eniser2025translatingrealworldcodellms}. Recent studies have demonstrated encouraging functional correctness for C-to-Rust translation~\cite{10.1145/3786180.3788317,wang2026evoc2rustskeletonguidedframeworkprojectlevel,yang2024vertverifiedequivalentrust,eniser2025translatingrealworldcodellms}. However, nearly all existing approaches adopt the same translation paradigm: the LLM receives the C source code and is asked to generate the corresponding Rust implementation. While effective for preserving program behavior, this \emph{code-only} setting also anchors the model to the original implementation. Consequently, the generated Rust often inherits C-specific implementation choices, including low-level control flow, data representations, and programming patterns, rather than taking advantage of Rust's higher-level abstractions and safety mechanisms.

This observation raises a question: can we provide the LLM with a representation that emphasizes the program's intended behavior rather than its implementation? In this work, we investigate \emph{Natural Language Specifications}~\cite{nitin2024spectra},  which we define as structured natural-language descriptions of program behavior, including inputs, outputs, preconditions, postconditions, invariants, error handling, functional description, state transitions and non-functional constraints. These specifications are generated from the original C source code, with one specification per program. Unlike formal specifications, they are expressed in natural language and do not provide mathematical guarantees of correctness or completeness. By emphasizing what a program does rather than how it is implemented, they may allow LLMs to produce more idiomatic Rust without being constrained by C-specific implementation details.

Recent work has begun exploring intermediate representations for LLM-based code translation. Dipto et al.~\cite{dipto2026algorithmbasedpipelinereliableintentpreserving}, for example, generate step-by-step natural-language descriptions before translation, while Nitin et al.~\cite{nitin2024spectra} use multi-modal specifications together with the source code. These studies demonstrate that richer intermediate representations can improve translation quality, but important questions remain unanswered. Existing natural-language intermediates primarily describe \emph{how} a program operates rather than \emph{what} behavior it guarantees. Likewise, natural language specifications have only been evaluated alongside the original source code, specifications that fail consistency checks are typically discarded rather than improved, and evaluations focus almost exclusively on functional correctness. As a result, it remains unclear whether test-validated specifications can serve as an effective standalone intermediate representation that improves not only correctness, but also maintainability, idiomaticity, and safety.

To answer these questions, we conduct a systematic comparison of three translation settings: \emph{code-only}, \emph{specification-only}, and \emph{code-and-specification}. We also study the effect of adding an iterative test-based validation and refinement step for specifications. Given a specification, we use an LLM to regenerate a C program and check its behavior against the available test cases. If the regenerated program fails, we use the observed failures to refine the specification. We call a specification test-validated when the regenerated program passes the available tests; this provides evidence of agreement on the tested behavior, rather than a guarantee of complete semantic equivalence. Using 300 C programs from IBM CodeNet~\cite{nitin2024spectra}, we evaluate GPT-5.2 and GPT-4o-mini across all three settings, measuring not only functional correctness but also maintainability, complexity, idiomaticity, unsafe construct usage, and translation cost.

Our results reveal a clear trade-off between correctness and code quality. Code-only translation achieves the highest test pass rate (98.6\%) but often mirrors C's implementation choices. In contrast, specification-only translation produces the most idiomatic Rust, reducing Clippy style warnings by 2.4$\times$ and unsafe blocks by approximately 6$\times$, while improving maintainability and reducing code complexity. Its main limitation is functional correctness: using specifications without validation yields test pass rates between 69\% and 86\%, depending on the experimental configuration. With test-based validation and refinement, the GPT-5.2 setting achieves a 95.3\% test pass rate. Finally, combining source code with test-validated specifications process provides the best overall balance, achieving a 98.3\% test pass rate while improving measured code quality relative to code-only translation.

Our main contributions are twofold: (i) a systematic comparison of code-only, specification-only, and code-and-specification translation, evaluating not only functional correctness but also maintainability, complexity, idiomaticity, unsafe construct usage, and cost; and (ii) an empirical evaluation of how test-based specification validation and refinement affect translation correctness and code quality, revealing trade-offs among specification quality, translation fidelity, code quality, and computational cost.

\section{Related Work}
\label{sec:related_work}


\textbf{Repository-Level Decomposition and Metadata-Driven Translation.} A common strategy is to decompose C programs into functional units and translate them incrementally using structural information such as function signatures, type definitions, and call graphs~\cite{shiraishi2024smartc2rust,wang2026evoc2rustskeletonguidedframeworkprojectlevel}. Eniser et al.~\cite{eniser2025translatingrealworldcodellms} take a different angle, focusing on functional correctness through differential fuzzing and structured prompting, though their success rates range from 20–60\% and drop sharply for programs over 100 lines. Tadesse et al.~\cite{tadesse2026} compare three transpilers against human-written Rust across 18 quality categories, finding that reducing unsafe constructs often introduces other problems like runtime panics or redundant code, a trade-off that correctness-only evaluation simply does not capture. None of these approaches use natural language specifications or isolate a specification-only setting; they remain anchored to syntactic and structural artifacts.

\textbf{Specifications and Program Properties as Translation Guides.}
Shetty et al.~\cite{shetty2024syzygy} extract program properties such as nullability, aliasing, and type information through dynamic analysis and use them to guide translation. These properties are used as-is, with no check that they are correct or complete, and they cover only a narrow slice of what a full specification can express. Nitin et al.~\cite{nitin2024spectra} take a broader view, showing that static behavioral specifications, input-output examples, and natural language descriptions each contribute something different, and that combining them with source code improves translation effectiveness. 
However, three important questions remain unanswered by both studies. First, neither ever tests specifications on their own without source code alongside them. Second, when a specification fails a consistency check, Nitin et al.~\cite{nitin2024spectra} discard it, and Shetty et al.~\cite{shetty2024syzygy} do not check at all. Neither tries to fix a specification that turns out to be wrong. Third, both studies measure only whether translations pass test cases, with no attention to whether the resulting code is maintainable, idiomatic, or safe. We address these three limitations in our work.

\textbf{Intermediate Representations and Prompt-Centric Refinement.} A third line of work asks whether giving the model a richer intermediate step can improve translation. Tai et al.~\cite{tai2025nl} have the model first summarize the source program in natural language before translating, gaining up to 13.8\% accuracy improvement on CodeNet and AVATAR ~\cite{ahmad2023avatar}. Dipto et al.~\cite{dipto2026algorithmbasedpipelinereliableintentpreserving} take a similar approach with a structured step-by-step description of control flow and I/O, improving accuracy by 10.8\%. Macedo et al.~\cite{macedo2024exploring} find that between 26.4\% and 73.7\% of translations require post-processing because model outputs mix code with natural language, while Ye et al.~\cite{ye2025prompt} improve results through iterative prompt refinement, achieving gains of up to 13.68\% on Java-to-Python translation.
The intermediates used by Tai et al.~\cite{tai2025nl} and Dipto et al.~\cite{dipto2026algorithmbasedpipelinereliableintentpreserving} describe how a program works rather than what it guarantees, and neither is checked against the source program before being used. The prompt-centric approaches tend to be problem-specific and do not generalize well across diverse inputs. Rather than refining the prompt, we refine the intermediate artifact itself, improving the specification through test-based validation until it accurately captures the program's behavior.

\section{Study Design}
\label{sec:study_design}

The goal of this study is to \textit{investigate whether natural language specifications can serve as an effective intermediate representation for LLM-based C-to-Rust translation}. Specifically, we examine whether reasoning from a program's intended behavior, rather than solely from its implementation, leads to translations that are not only functionally correct but also more maintainable, idiomatic, and safe. Our study is guided by the following research questions (RQs):

\noindent
\textbf{RQ1: How does test-based validation and refinement of specifications affect specification-based translation?} Specification-based translation can only be as correct as the specification it starts from. This RQ compares unvalidated and test-validated specifications to measure how much checking and refining a specification before translation changes translation correctness. We assess whether it carries enough information to reproduce the program’s behavior.

\noindent
\textbf{RQ2: What is the impact of specifications on the trade-off between translation correctness and code quality?} Code-only translation gives the LLM access to the implementation, so we expect it to achieve the highest correctness. This RQ asks whether that correctness advantage comes at the cost of code quality, and whether specification-based translation can close that quality gap without sacrificing too much correctness.

\noindent
\textbf{RQ3: To what extent do specifications encourage implementation choices that are less constrained by the original source implementation?} This RQ looks past aggregate metrics to ask what is actually happening at the code level: when an LLM translates from a specification instead of code, does it produce implementations that are free from C's structural choices, or does it still end up mirroring C's structure?

\noindent
\textbf{RQ4: What are the cost–quality trade-offs of specification-based translation pipelines?} Generating, validating, and refining specifications all require additional LLM calls beyond translation itself. This RQ asks whether the quality gains from specification-based translation justify their added cost.
\vspace{2mm}

\begin{figure}[!tp]
    \centering
    \includegraphics[width=\columnwidth]{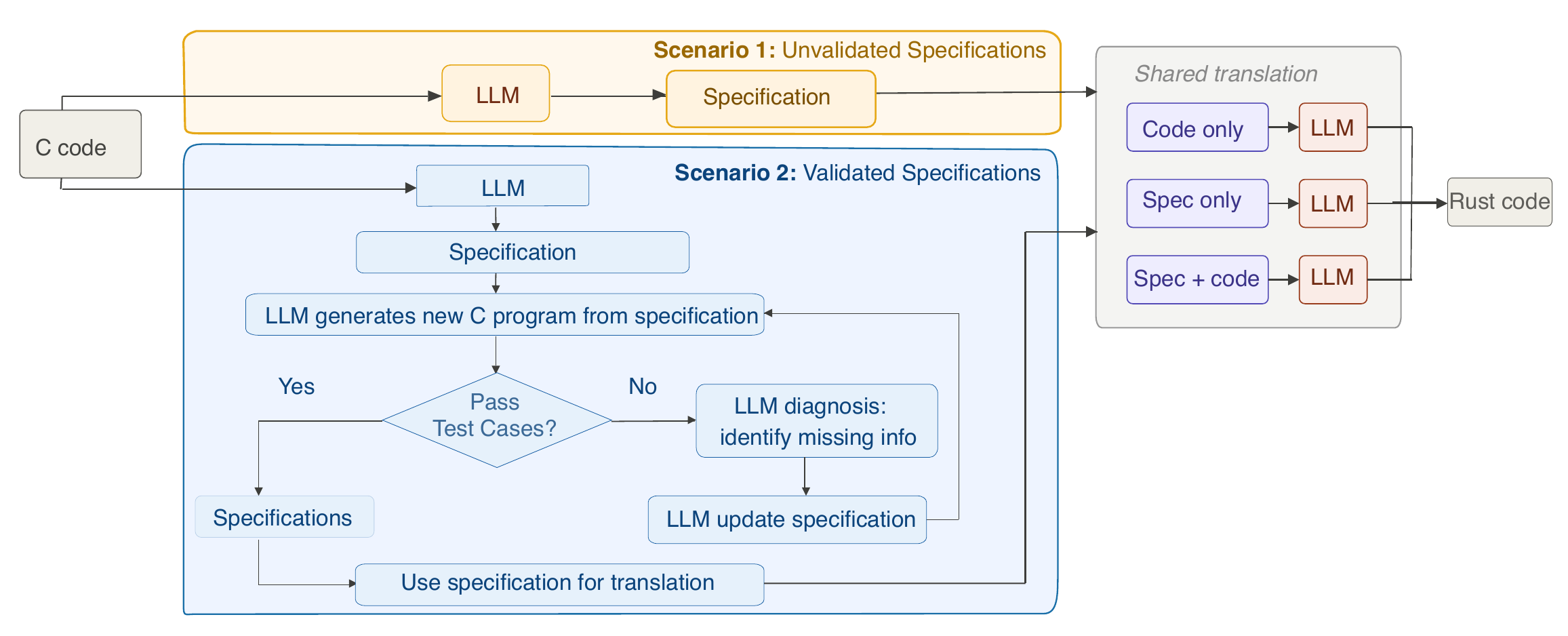}
    \caption{Overview of Translation Approach.}
    \label{fig:process_overview}
\end{figure}

Figure~\ref{fig:process_overview} provides a high-level overview of our study. Given a source program, we first generate a natural-language specification using an LLM, then translate the source program into the target language across three settings (i.e., code-only, specification-only, and code-and-specification). These three settings allow us to systematically isolate and measure the contribution of specifications to translation quality. In the following subsections, we detail our dataset selection criteria, LLM configuration, and specification generation strategy.

\subsection{Dataset}
We leverage the IBM CodeNet~\cite{puri2021codenetlargescaleaicode} dataset as the primary corpus for this study. CodeNet comprises over 4,000 programming problems and test cases, in multiple programming languages, collected from competitive programming platforms. 
This dataset has been adopted in several studies~\cite{ye2025prompt, macedo2024exploring, macedo2025intertrans, ding2024traced}. To ensure comparability with prior work, we adopt the subset of 300 function-level code submissions written in C sampled by Nitin et al.~\cite{nitin2024spectra}. 
Thus, our study focuses exclusively on the C-to-Rust translation setting. This choice is also motivated by the growing interest in migrating legacy C systems to Rust~\cite{ou2025rustrepotrans, emre2021translating}. 

\subsection{LLM Selection}

We select two LLMs from \textit{OpenAI}, namely \textit{GPT-5.2} (latest model as the time of conducting this experiment) and \textit{GPT-4o-mini} due to the wide adoption of these family of models in recent works~\cite{nitin2024spectra,C2SaferRust2025}. These models were chosen to balance high-capability reasoning models with more lightweight, cost-efficient alternatives, enabling a comprehensive evaluation of different performance and efficiency trade-offs. 
We set the temperature to 0 to ensure deterministic output, while keeping all other model parameters at their defaults.

\subsection{Specification Generation, Validation, and Refinement}
To derive natural language specifications, we prompt an LLM to generate structured specifications directly from the given code. Our specification format builds on the specifications used by Shetty et al.~\cite{shetty2024syzygy} and Nitin et al.~\cite{nitin2024spectra}, extending their structure with additional sections, namely invariants, error handling, state transitions, and non-functional constraints, and with explicit instructions asking the LLM to separate what a program guarantees from how it achieves it. The specification describes the observable behavior of a self-contained program. It does not describe memory-level properties such as ownership, aliasing, or resource lifetimes, nor does it capture cross-function contracts, shared global state, or concurrency. These properties are central to repository-scale C-to-Rust migration but largely absent from the self-contained CodeNet programs we study; extending the format to express them is left to future work. Listing~\ref{fig:spec_prompt} presents the prompt template for the specification generation. 

Using this prompt, we consider two scenarios. In the first scenario, we use the LLM to generate the specification of a given C code and use this specification for translation to Rust in future steps without validation to check whether the specification is correct or not. In the second scenario, we incorporate a \textit{test-based validation step} to assess the correctness of the generated specification before we utilize it for translation. Concretely, after generating the specification using the same prompt in Listing~\ref{fig:spec_prompt}, we prompt the LLM in a separate request to synthesize an equivalent C program from the generated specification. The resulting program is then executed against the original test inputs. 

If the synthesized program passes all test cases, we consider the specification test-validated: it contains enough information to reproduce the program’s behavior on the available tests. This does not establish that the specification is complete or correct beyond the tested behaviors; incorrect or extraneous statements that the tests do not exercise may remain undetected. If the synthesized program fails the test cases, we apply a specification diagnostic step consisting of two prompts, which we have made available in our supplementary material~\cite{supplementary}. In this step, we prompt the LLM to identify discrepancies between the generated specification and the original C program, and between expected and actual outputs. The resulting diagnosis is then used, together with the original source code and the current specification, to \textit{refine} the specification using a subsequent LLM prompt. The test cases used for validation are the same ones later used to evaluate the Rust translations. To limit how directly test information enters the specification, the refinement prompt does not receive the test cases themselves, only the natural-language diagnosis; the specification is thus revised based on a description of the behavioral gap rather than on specific input--output pairs. This process is repeated until the program regenerated from the specification passes all available tests. If, after $k$ attempts ($k=3$ in our study), the synthesized C program still fails, we retain the most recently generated specification for that problem as the final output. This process is shown in Figure~\ref{fig:process_overview}. Throughout the paper, we call specifications produced by this process \textit{test-validated}, and specifications used directly after generation, without this check, \textit{unvalidated}. For those where the regenerated program still failed after $k=3$ attempts; we retain the last specification for these programs and include them in the test-validated set.


\begin{listing}[!tp]
\centering

\begin{tcolorbox}[
  colframe=gray,
  colback=white,
  boxrule=0.5pt,
  arc=0.3em,
  boxsep=1mm,
  left=1mm,
  right=1mm,
  top=1mm,
  bottom=1mm,
  fontupper=\ttfamily\scriptsize
]
\begingroup
\obeylines
\obeyspaces

You are an expert senior software engineer.
Given source code, generate a precise, structured behavioral specification that states WHAT the code guarantees, not HOW it achieves it.

Output must use exactly these sections and headings:
1. Functional Description
2. Inputs
3. Outputs
4. Preconditions
5. Postconditions
6. Invariants
7. Error Handling / Exceptional Behavior
8. State Transitions (if applicable)
9. Non-functional Constraints (if inferable)

Content requirements:
- Be concise and contract-focused; omit incidental or low-level details.
- Do not reference language/library APIs, format strings, or identifiers (e.g., no mentions of scanf, printf, fgets, variable or function names, loop types, or recursion).
- Describe I/O abstractly (e.g., "reads values from standard input" or "returns a value"); never cite exact parsing patterns or format specifiers.
- Do not restate implementation structure or control flow. Focus on observable behavior, data constraints, and guarantees.
- If behavior is ambiguous, state explicit assumptions. If something cannot be deduced, write "Not inferable from code."
- Avoid extreme assumptions. Capture only behavior implied by the code’s observable contract.
- If the program is an executable, treat printing/writing as side effects in Outputs.
- No code, no pseudocode, no preamble or epilogue text. Use only the nine sections above. 
- Return only the nine sections with the exact headings and numbering specified. Do not include any other text, examples, or code.
Now generate the specification for the code below:

{<code>}

\endgroup
\end{tcolorbox}
\caption{Prompt template used for specification generation} 
\label{fig:spec_prompt}
\end{listing}

\subsection{Translation to Rust}
\label{sec:rust_code_generation}


We evaluate C-to-Rust translation under three variants: \textit{code-only}, \textit{specification-only}, and \textit{code-and-specification}. Each variant allows up to $k=3$ repair attempts, a limit we chose based on observed diminishing returns in success rates beyond this point.
When an initial translation fails, the LLM is iteratively prompted to repair it using an execution report derived from the previous attempt. This report may include Compilation Errors (CE), Runtime Errors (RE), Timeout Failures (TO), or mismatches between expected and actual outputs (test failures). Repair continues until a correct translation is produced or all attempts are exhausted. The prompts used at this stage are provided in our supplementary material~\cite{supplementary}.

\subsection{Internal Code Quality Measurements}
To measure the internal quality of the translated Rust code, we use \texttt{rust-code-analysis}~\cite{rust-code-analysis}, a static analysis tool developed by Mozilla for computing source code metrics, which has also been adopted in prior work on Rust code quality evaluation~\cite{dehghan2025translatinglargescalecrepositories}. We use it to compute maintainability (Maintainability Index, SEI variant), complexity (Cyclomatic and Cognitive Complexity), and size and effort metrics (Halstead Volume and Effort, SLOC, PLOC, LLOC, comment density, and blank lines).

\subsection{Idiomaticity Measurement}
To measure the idiomaticity of the generated Rust code, we use Clippy ~\cite{clippytool}, the official Rust linter. We exclude lints from the nursery and restriction categories, as nursery lints are still under development and may produce false positives, and Clippy's own documentation advises against enabling restriction lints by default due to their subjective nature. We also exclude pedantic lints since they are disabled by default, are opinionated, and may produce false positives. Finally, we exclude the cargo category as those lints target Cargo.toml configuration files rather than the source code itself.

We normalize raw warning counts by the original C lines of code (LOC), putting all three settings on a common baseline. 
This allows comparing how many issues each setting introduces per unit of input program, a consistent approach used in previous papers~\cite{dehghan2025translatinglargescalecrepositories}. 

In addition to idiomaticity, we also measure the use of unsafe constructs in the generated Rust code. A key motivation for migrating C code to Rust is to leverage Rust's safety guarantees, so we want to know whether the translated code actually avoids unsafe constructs or simply replicates C-style patterns. We measure three things: the number of unsafe blocks, the number of lines covered by unsafe blocks, and the number of type casts. Unsafe blocks are counted by traversing the High-Level Intermediate Representation (HIR) and identifying blocks marked with the unsafe keyword. For unsafe lines, we collect the set of source lines spanned by those blocks to avoid double-counting lines that fall under nested unsafe blocks. Type casts are counted by identifying explicit cast expressions in the HIR. We use a custom static analyzer built on top of the Rust compiler internals to collect these metrics. We have included this analyzer in our supplementary material~\cite{supplementary}. As with Clippy warnings, we normalize the raw counts by the C LOC.

\section{Results}
\label{sec:results}
This section presents the results and analysis from our study and also answers to our RQs. To support our analysis with code snippets, we developed a website\footnote{\url{http://74.208.64.171:8000/translations}} to make the translated code and generated specification artifacts easily accessible. 

\subsection{RQ1: How does test-based validation and refinement of specifications affect specification-based translation?}
\label{sec:RQ1}


\paragraph{\textbf{Performance of GPT-4o-mini with Unvalidated Specifications}}

We first evaluate the translation performance of GPT-4o-mini with unvalidated specifications. As shown in Table~\ref{fig:gpt_4o_mini_status}, the model demonstrates limited effectiveness, achieving a maximum of 147 successful translations in the code-and-specification setting at $k=3$, and 138 and 90 at the code-only and specification-only settings, respectively. It is worth noting that while the specification-only variant produced fewer compilation errors, it did not improve correctness (pass rate); instead, it led to higher test failure rates. With this observation, we replaced GPT-4o-mini with GPT-5.2 as the model responsible for translation to target language (Rust) while retaining the same specifications generated by  GPT-4o-mini. The result of this model transition is shown in Table~\ref{fig:gpt_5_2_using_gpt_4o_mini_status}. This operation yielded substantial improvements, with code and specification increasing from 147 to 295, code-only successes increasing from 138 to 295,  specification only increasing from 90 to 221 and  at $k=3$. This improvement is attributable to the stronger reasoning and code synthesis capabilities of GPT-5.2. Nevertheless, the specification-only variant still remains a limiting factor, with 75 failures at $k=3$, suggesting that specification quality constrains performance independently of model capability.

\begin{table}[t]
\caption{Translations with GPT-4o-mini using Specifications Generated by GPT-4o-mini}
\label{fig:gpt_4o_mini_status}
\centering
\scriptsize
\begin{tabular}{@{}llrrrrrr@{}}
\toprule
\textbf{Profile} & \textbf{k} & \textbf{Total} & \textbf{Pass (Cum.)} &
\textbf{Fail} & \textbf{RE} & \textbf{TO} &
\textbf{CE} \\
\midrule
\multirow{3}{*}{Code+Spec}
 & 1 & 300 & 112 (\textbf{112}) & 20  & 46 & 0 & 122 \\
 & 2 & 188 & 26  (\textbf{138}) & 24  & 49 & 1 & 88  \\
 & 3 & 162 & 9   (\textbf{147}) & 25  & 45 & 0 & 83  \\
\midrule
\multirow{3}{*}{Code Only}
 & 1 & 300 & 103 (\textbf{103}) & 8   & 41 & 2 & 146 \\
 & 2 & 197 & 25  (\textbf{128}) & 13  & 60 & 3 & 96  \\
 & 3 & 172 & 10  (\textbf{138}) & 19  & 42 & 4 & 97  \\
\midrule
\multirow{3}{*}{Spec Only}
 & 1 & 300 & 45  (\textbf{45})  & 125 & 29 & 2 & 99  \\
 & 2 & 256 & 23  (\textbf{68})  & 175 & 25 & 2 & 31  \\
 & 3 & 237 & 22  (\textbf{90})  & 156 & 21 & 2 & 36  \\
\bottomrule
\end{tabular}
\end{table}

\begin{table}[t]
\caption{Translations with GPT-5.2 using Specifications Generated by GPT-4o-mini}
\label{fig:gpt_5_2_using_gpt_4o_mini_status}
\centering
\scriptsize
\begin{tabular}{@{}llrrrrrr@{}}
\toprule
\textbf{Profile} & \textbf{k} & \textbf{Total} & \textbf{Pass (Cum.)} &
\textbf{Fail} & \textbf{RE} & \textbf{TO} &
\textbf{CE} \\
\midrule
\multirow{3}{*}{Code+Spec}
 & 1 & 300 & 279 (\textbf{279}) & 9   & 7  & 0 & 5 \\
 & 2 & 21  & 16  (\textbf{295}) & 3   & 1  & 0 & 1 \\
 & 3 & 5   & 0   (\textbf{295}) & 4   & 1  & 0 & 0 \\
\midrule
\multirow{3}{*}{Code Only}
 & 1 & 300 & 280 (\textbf{280}) & 6   & 7  & 1 & 6 \\
 & 2 & 20  & 11  (\textbf{291}) & 7   & 2  & 0 & 0 \\
 & 3 & 9   & 4   (\textbf{295}) & 4   & 1  & 0 & 0 \\
\midrule
\multirow{3}{*}{Spec Only}
 & 1 & 300 & 111 (\textbf{111}) & 160 & 26 & 0 & 3 \\
 & 2 & 189 & 74  (\textbf{185}) & 110 & 5  & 0 & 0 \\
 & 3 & 115 & 36  (\textbf{221}) & 75  & 2  & 1 & 1 \\
\bottomrule
\end{tabular}
\end{table}


\begin{table}[t]
\caption{Translations with GPT-5.2 using Specifications Generated by GPT-5.2}
\label{fig:gpt_5_2_using_gpt_5_2_specs_status}
\centering
\scriptsize
\begin{tabular}{@{}llrrrrrr@{}}
\toprule
\textbf{Profile} & \textbf{k} & \textbf{Total} & \textbf{Pass (Cum.)} &
\textbf{Fail} & \textbf{RE} & \textbf{TO} &
\textbf{CE} \\
\midrule
\multirow{3}{*}{Code+Spec}
 & 1 & 300 & 266 (\textbf{266}) & 10  & 5  & 18 & 1 \\
 & 2 & 34  & 12  (\textbf{278}) & 3  & 2  & 17 & 0 \\
 & 3 & 22  & 10  (\textbf{288}) & 2  & 3  & 7  & 0 \\
\midrule
\multirow{3}{*}{Code Only}
 & 1 & 300 & 275 (\textbf{275}) & 5  & 6  & 9  & 5 \\
 & 2 & 25  & 13  (\textbf{288}) & 6  & 2  & 3  & 1 \\
 & 3 & 12  & 3   (\textbf{291}) & 4  & 2  & 3  & 0 \\
\midrule
\multirow{3}{*}{Spec Only}
 & 1 & 300 & 216 (\textbf{216}) & 64 & 13 & 4  & 3 \\
 & 2 & 84  & 34  (\textbf{250}) & 44 & 6  & 0  & 0 \\
 & 3 & 50  & 9   (\textbf{259}) & 36 & 3  & 0  & 2 \\
\bottomrule
\end{tabular}
\end{table}

\paragraph{\textbf{Performance of GPT-5.2 with Unvalidated Specifications}}
Based on the previous observations, we conducted an additional experiment in which \textit{GPT-5.2} is used both to generate specifications and to perform the subsequent code translation. The results of this configuration are presented in Table~\ref{fig:gpt_5_2_using_gpt_5_2_specs_status}. 
Overall, we observe trends consistent with the earlier setup. In particular, the code-and-specification and code-only variants achieve higher test pass rates (288 and 291, respectively) compared to the specification-only variant (259). The specification-only setting continues to exhibit relatively high failure rates, with 62 failures at \(k=1\) and 34 failures at \(k=3\).


Given the persistent pattern of high test failure rates in specification-only translation, a detailed analysis of selected cases revealed that specification incompleteness is a key limiting factor. Consider the example in Figure~\ref{fig:spec-validation-example}, which shows a cube root computation using Newton-Raphson iteration. The unvalidated specification (see in~\cite{p00080}) omits the initial guess, causing the LLM to default to $r=1.0$ instead of $p0 = q/2$, and describes the iteration abstractly, leading the LLM to invert the loop structure relative to the original \texttt{do-while}. The test-validated specification~\cite{p00080} explicitly corrects both omissions, yielding a correct translation. To address such cases systematically, we introduce a test-based validation and refinement step to improve specification fidelity.


\begin{figure*}[t]
\centering
\begin{subfigure}[t]{\linewidth}
  \caption*{\textbf{(a) Original C source}}
  \vspace{2pt}
  \begin{lstlisting}[style=cstyle]
while (scanf("%d", &q) != EOF && (*@\colorbox{hlamber}{\texttt{q != -1}}@*)) {
    (*@\colorbox{hlamber}{\texttt{px = (double)q / 2;}}@*)          // initial guess = q/2
    do {
        x = px - (px*px*px - q) / (3*px*px);  // Newton step
        px = x;
    } while (fabs(x*x*x - q) >= EPS * q);     // stop when |x^3-q| < eps*q
    printf("%.6lf\n", x);                      // print final x
}
  \end{lstlisting}
\end{subfigure}
 
\vspace{6pt}

\begin{subfigure}[t]{\linewidth}
  \caption*{\textbf{(b) Key differences between unvalidated and test-validated specification}}
  \vspace{2pt}
  \begin{tabularx}{\linewidth}{>{\bfseries\scriptsize}p{2.1cm}
                                >{\scriptsize}X
                                >{\scriptsize}X}
    \toprule
    \textbf{Aspect}
      & \cellcolor{diffold}\textcolor{diffoldt}{\textbf{Unvalidated specification}}
      & \cellcolor{diffnew}\textcolor{diffnewt}{\textbf{Validated specification}} \\
    \midrule
    Sentinel value
      & \cellcolor{diffold}\textcolor{diffoldt}{Described as an unspecified ``designated termination value''}
      & \cellcolor{diffnew}\textcolor{diffnewt}{Explicitly \texttt{-1}; sentinel is read but not processed} \\
    Initial guess
      & \cellcolor{diffold}\textcolor{diffoldt}{Omitted entirely}
      & \cellcolor{diffnew}\textcolor{diffnewt}{$p_0 = q/2$ (as a \texttt{double}) — stated explicitly} \\
    Newton step
      & \cellcolor{diffold}\textcolor{diffoldt}{Described only as an ``iterative method''}
      & \cellcolor{diffnew}\textcolor{diffnewt}{Full formula: $x = p - (p^3 - q)\,/\,(3p^2)$, then $p \leftarrow x$} \\
    Loop semantics
      & \cellcolor{diffold}\textcolor{diffoldt}{Stopping condition stated; loop form unspecified}
      & \cellcolor{diffnew}\textcolor{diffnewt}{Do-while semantics explicit: at least one iteration guaranteed} \\
    Output value
      & \cellcolor{diffold}\textcolor{diffoldt}{``Approximation satisfying the inequality''}
      & \cellcolor{diffnew}\textcolor{diffnewt}{``The final $x$ when the loop terminates'' — not an arbitrary satisfying value} \\
    \bottomrule
  \end{tabularx}
\end{subfigure}
 
\vspace{6pt}

\begin{subfigure}[t]{0.48\linewidth}
  \caption*{\textbf{(c\textsubscript{1})} Rust from \textit{unvalidated} spec
            \hfill\colorbox{diffold}{\textcolor{diffoldt}{\scriptsize incorrect}}}
  \vspace{2pt}
  \begin{lstlisting}[style=ruststyle]
// initial guess r = 1  <- not in spec
let mut r = (*@\colorbox{hlred}{\texttt{1.0f64}}@*);
let tol = 1e-5f64 * q;
for _ in 0..1_000_000 {
    if (r*r*r - q).abs() < tol { break; }
    let r2 = r * r;
    // algebraically different Newton form
    (*@\colorbox{hlred}{\texttt{r = (2.0*r + q/r2) / 3.0;}}@*)
}
println!("{:.6}", r);
  \end{lstlisting}
\end{subfigure}
\hfill
\begin{subfigure}[t]{0.48\linewidth}
  \caption*{\textbf{(c\textsubscript{2})} Rust from \textit{test-validated} spec
            \hfill\colorbox{diffnew}{\textcolor{diffnewt}{\scriptsize correct}}}
  \vspace{2pt}
  \begin{lstlisting}[style=ruststyle]
// p_0 = q/2, as specified
let mut p: f64 = (*@\colorbox{hlgreen}{\texttt{q / 2.0}}@*);
let x_final: f64 = loop {
    (*@\colorbox{hlgreen}{\texttt{let x = p - (p*p*p - q) / (3.0*p*p);}}@*)
    p = x;
    if (x*x*x - q).abs() < eps * q {
        break x;  // final x, as specified
    }
};
println!("{:.6}", x_final);
  \end{lstlisting}
\end{subfigure}

\caption{
  (a) Original C source. Code comments in all panels are annotations added by the authors for readability; they were not part of the input given to the LLM, nor part of the LLM output. Amber highlights mark the two behavioral details
  most critical to faithful translation.
  (b) Differences between the unvalidated and test-validated specifications, summarized manually by the authors; the "Aspect" labels are our own categories and are not part of either specification.
  (c\textsubscript{1}) and~(c\textsubscript{2}) show the resulting Rust
  translations: the unvalidated specification causes the LLM to choose an initial guess of
  $r{=}1$ and a different algebraic form of Newton's method, whereas the test-validated specification yields a translation that reproduces the program's observable behavior.
  \colorbox{hlred}{\phantom{x}}~Red highlights indicate deviations;
  \colorbox{hlgreen}{\phantom{x}}~green highlights indicate correct correspondence.}
\label{fig:spec-validation-example}
 
\end{figure*}

\begin{table}[t]
\caption{Translations with GPT-5.2 using test-validated specifications generated by GPT-5.2}
\label{fig:gpt_5_2_using_gpt_5_2_validated_specs_status}
\centering
\scriptsize
\begin{tabular}{@{}llrrrrrr@{}}
\toprule
\textbf{Profile} & \textbf{k} & \textbf{Total} & \textbf{Pass (Cum.)} &
\textbf{Fail} & \textbf{RE} & \textbf{TO} &
\textbf{CE} \\
\midrule
\multirow{3}{*}{Code+Spec}
 & 1 & 300 & 281 (\textbf{281}) & 9  & 8  & 0 & 2 \\
 & 2 & 21  & 13  (\textbf{294}) & 7  & 1  & 0 & 0 \\
 & 3 & 8   & 1   (\textbf{295}) & 6  & 1  & 0 & 0 \\
\midrule
\multirow{3}{*}{Code Only}
 & 1 & 300 & 282 (\textbf{282}) & 7  & 5  & 2 & 4 \\
 & 2 & 18  & 12  (\textbf{294}) & 6  & 0  & 0 & 0 \\
 & 3 & 6   & 2   (\textbf{296}) & 4  & 0  & 0 & 0 \\
\midrule
\multirow{3}{*}{Spec Only}
 & 1 & 300 & 274 (\textbf{274}) & 15 & 7  & 0 & 4 \\
 & 2 & 28  & 10  (\textbf{284}) & 15 & 1  & 1 & 1 \\
 & 3 & 18  & 2   (\textbf{286}) & 15 & 0  & 1 & 0 \\
\bottomrule
\end{tabular}
\end{table}


Based on our previous observations of LLM performance with GPT-4o-mini and GPT-5.2 on unvalidated specifications, we proceeded with the rest of our experiments using GPT-5.2.

\paragraph{\textbf{Performance of GPT-5.2 with Test-validated Specifications}}
Table~\ref{fig:gpt_5_2_using_gpt_5_2_validated_specs_status} presents the results using test-validated specifications for translation. The results indicate that specification refinement leads to substantial improvements, particularly for the \textit{specification-only} variant. 
Specifically, when compared to the non-validated setting (Table~\ref{fig:gpt_5_2_using_gpt_5_2_specs_status}) for specification-only translation, the test failure rate decreases significantly from 62 to 13 at \(k=1\), and the number of successfully translated code increases from 217 to 274 ($\approx$26\% increase). After three attempts (\(k=3\)), the failure rate further reduces from 34 to 11 for the specification-only variant. Correspondingly, the number of successfully translated programs increases from 259 to 286.

In the unvalidated setting, overly abstract specifications lead to plausible but incorrect translations. Refinement addresses this by adding the missing behavioral details, and some of these details are implementation-level. In the example in Figure~\ref{fig:spec-validation-example}, the refined specification states the initial guess and that the update precedes the stopping check, because the printed value depends on both. This is a limitation of our approach: refinement can move a specification from describing what a program guarantees toward describing how it computes it, which may in turn steer the translation back toward the original implementation. We did not measure how often this occurs. However, translations from test-validated specifications still replaced C’s algorithms or data structures in most programs (Section~\ref{sec:translation-differences}), suggesting that refinement did not reintroduce C’s implementation wholesale.

\paragraph{\textbf{Idiomaticity of the Generated Code}}
Table~\ref{tab:clippy}  presents both the raw and normalized Clippy-based idiomaticity counts across the different translation settings. Across four of the five categories, specification-only translations produce fewer warnings than both code-only and code-and-specification, suggesting that specification tends to produce more idiomatic Rust. The one exception is the correctness category, where specification-only produces 6 warnings compared to 2 for code-only and 0 for code-and-specification.

A closer look at the individual warnings reveals that these correctness issues are not fundamental translation errors but artifacts of LLM-type reasoning without source code, manifesting as redundant guards that are logically vacuous in Rust's type system. Code-only warnings, by contrast, stem from structural mimicry of C code.
We note that the absolute numbers here are very small, at most 8 warnings across 900 programs, so we do not draw strong conclusions from this category alone. The overall picture from Table~\ref{tab:clippy} still favors specification-only and code-and-specification over code-only, particularly for the style and complexity categories which account for the large majority of warnings.

Table~\ref{tab:unsafe} reports unsafe construct usage. Specification-only translation produces far fewer unsafe blocks (14) and lines (14) than code-only (82 blocks, 342 lines), a $6\times$ and $24\times$ reduction, respectively, with code-and-specification in between (22 blocks, 48 lines). The same trend holds for type casts: 691 for specification-only versus 1,756 for code-only. This is consistent with the qualitative observation in RQ3 (Section~\ref{sec:translation-differences}): 
Without direct access to C code, the model is no longer anchored to C's low-level patterns and instead produces translations that adopt safe Rust idioms. Together, Tables~\ref{tab:clippy} and~\ref{tab:unsafe} present a consistent picture that specification-based translation produces more idiomatic and safer Rust meaningfully. 

To confirm that these differences are statistically meaningful, we applied the Friedman test~\cite{cad058f0-09f2-35e2-b8ad-9038985e0961} followed by post-hoc Wilcoxon signed-rank~\cite{c4091bd3-d888-3152-8886-c284bf66a93a} tests with Holm-Bonferroni~\cite{29def780-e117-38f0-8afb-edf384af3fad} correction across the 285 programs. These tests were chosen because Shapiro--Wilk~\cite{10.1093/biomet/52.3-4.591} normality tests indicated non-normal distributions (all $p < 0.05$), the paired nature of the data favors the Friedman test over the Kruskal--Wallis test, and the Holm--Bonferroni correction balances family-wise error control with sensitivity across three pairwise comparisons~\cite{29def780-e117-38f0-8afb-edf384af3fad}.
For total Clippy warnings ($\chi^2(2) = 119.12$, $p < 0.001$), specification-only produced significantly fewer warnings than both code-and-specification ($p < 0.001$, $r = 0.61$, large effect) and code-only ($p < 0.001$, $r = 0.74$, large effect). The style category, which accounts for the large majority of warnings, drove this result most strongly ($\chi^2(2) = 123.26$, $p < 0.001$), with code-only producing significantly more style warnings than both specification-only ($r = 0.73$, large effect) and code-and-specification ($r = 0.50$, large effect). Complexity warnings followed the same pattern ($\chi^2(2) = 17.36$, $p < 0.001$), with code-only producing significantly more than both specification-based settings. For unsafe blocks ($\chi^2(2) = 25.42$, $p < 0.001$), both specification-only and code-and-specification produced significantly fewer unsafe blocks than code-only ($r = 1.00$ and $r = 0.97$, respectively, both large effects), while the difference between the two specification-based settings was not statistically significant ($p = 0.059$). Type casts showed significant differences across all three settings ($\chi^2(2) = 141.66$, $p < 0.001$), with code-only producing the most ($r = 0.77$ vs.\ specification-only, $r = 0.75$ vs.\ code-and-specification) and specification-only the fewest.

\begin{table}[!tp]
\centering
\caption{Code size of Original C and Rust code}
\label{tab:tsi}
\begin{tabular}{lcc}
\toprule
\textbf{Translation} & \textbf{C LOC} & \textbf{Rust LOC}  \\
\midrule
specification-only    &18,527 &24,226  \\
code-only    &18,527 &27,842  \\
code-and-specification  &18,527 &23,407  \\
\bottomrule
\end{tabular}
\end{table}

\begin{table}[!tp]
\centering
\caption{Clippy warning counts (Raw, Normalized). Highlighted cells indicate the lowest normalized value per row.}
\label{tab:clippy}
\begin{tabular}{lp{2cm}p{2cm}p{1.5cm}}
\toprule
\textbf{Category} & \textbf{specification-only} & \textbf{code-only} & \textbf{code-and-specification} \\
\midrule
Correctness  & (6, 0.32) & (2, 0.11) & \cellcolor{bestcell}(0, 0.00) \\
Suspicious   & \cellcolor{bestcell}(0, 0.00) & \cellcolor{bestcell}(0, 0.00) & \cellcolor{bestcell}(0, 0.00) \\
Complexity   & \cellcolor{bestcell}(76, 4.10) & (186, 10.04) & (114, 6.15) \\
Performance  & \cellcolor{bestcell}(14, 0.76) & (30, 1.62) & (18, 0.97) \\
Style        & \cellcolor{bestcell}(456, 24.61) & (1084, 58.51) & (842, 45.45) \\
\midrule
\end{tabular}
\end{table}

\begin{table}[!tp]
\centering
\caption{Unsafe construct counts (Raw, Normalized). Highlighted cells indicate the lowest normalized value per row.}
\label{tab:unsafe}
\begin{tabular}{lp{2cm}p{2cm}p{1.5cm}}
\toprule
\textbf{Metric} & \textbf{specification-only} & \textbf{code-only} & \textbf{code-and-specification} \\
\midrule
Unsafe Blocks  & \cellcolor{bestcell}(14, 0.76) & (82, 4.43)          & (22, 1.19) \\
Unsafe Lines   & \cellcolor{bestcell}(14, 0.76) & (342, 18.46)         & (48, 2.59) \\
Type Casts     & \cellcolor{bestcell}(691, 37.30) & (1756, 94.78)      & (884, 47.71) \\
\bottomrule
\end{tabular}
\end{table}


\begin{tcolorbox}[
  title=\textbf{RQ1 Answer},
  fonttitle=\bfseries,
  colback=white,
  boxrule=0.5pt,
  arc=0.3em,
  boxsep=1mm,
  left=1mm,
  right=1mm,
  top=1mm,
  bottom=1mm,
]

Specification-only translation depends strongly on whether the specification passes test-based validation. Unvalidated specifications often omit behavioral details that the program’s observable behavior depends on, leading to pass rates around 86\% (GPT-5.2). Our iterative back-translation refinement closes this gap, raising the specification-only pass rate to 95.3\%.
\end{tcolorbox}

\subsection{RQ2: What is the impact of specifications on the trade-off between translation correctness and code quality?}
\label{sec:quality_metrics}

\paragraph{\textbf{Analysis of Code Quality Metrics Across Specification Sources}}
\label{sec:quality_metrics_unvalidated}
Table~\ref{tab:gpt52_comparison} presents a comparison of code quality metrics across three input configurations. To ensure that observed differences reflect genuine variation in code quality rather than differences in problem coverage, we restricted this evaluation to the intersection of problems successfully translated under all three input settings (code-only, specification-only, and code-and-specification) across the three experiment setups reported in Tables~\ref{fig:gpt_5_2_using_gpt_4o_mini_status}, \ref{fig:gpt_5_2_using_gpt_5_2_specs_status}, and \ref{fig:gpt_5_2_using_gpt_5_2_validated_specs_status}. This intersection yielded a balanced subset of 193 problems on which all reported metrics were computed.

The results carry a clear message for practitioners: the context you give the model has effect on code quality attributes. Translations produced from \textit{specification-only} prompts were consistently the most readable and maintainable across all three experiment setups. They used roughly 30--35\% fewer logical lines of code than code-only translations (e.g., 32.90 vs.\ 46.90 LLOC in the first configuration) and generated 44--55\% more comments (11.39 vs.\ 7.92 CLOC), which may be a meaningful difference for developers inheriting such code. Halstead Effort scores, which approximate the cognitive work required to understand a function, were nearly 45\% lower for specification-only translations than for code-only ones (${\sim}89{,}000$ vs.\ ${\sim}161{,}000$--$167{,}000$). The Maintainability Index (SEI) reinforces this picture: code-only translations hovered around 38--39, falling well below the threshold of 65 associated with low maintainability~\cite{herivcko2023exploring}, while specification-only 
translations reached 45--49, narrowing that gap considerably.

The one exception is cyclomatic complexity, where \textit{code-and-specification} consistently won (13.44--14.11 vs.\ 15.04--16.53 for specification-only), suggesting that access to the original source helps the model preserve the branching structure of the reference implementation which is useful when control-flow conformance matters. \textit{Code-only} prompting ranked the least on every other dimension, producing the largest, least commented, and most cognitively demanding output regardless of which model generated the specifications.


It is also important to note that, high structural quality does not guarantee functional correctness. Specifications generated by GPT-4o-mini produce more maintainable code but are associated with higher test failure rates when used in isolation, suggesting they omit critical behavioral details. Concise specifications may be structurally sufficient yet semantically underspecified, yielding implementations that are clean but incomplete. The validation process addresses this directly: by refining natural language descriptions, it improves both the correctness and consistency of generated code, bridging the gap between structural tidiness and functional fidelity.

\begin{table*}[!t]
\centering
\resizebox{\textwidth}{!}{
\begin{tabular}{ll|ccc|ccc|ccc}
\hline
\textbf{Metric} & \textbf{Pref.}
& \multicolumn{3}{c|}{\textbf{GPT-5.2 + unvalidated specs (GPT-4o-mini)}}
& \multicolumn{3}{c|}{\textbf{GPT-5.2 + test-validated specs (GPT-5.2)}}
& \multicolumn{3}{c}{\textbf{GPT-5.2 + unvalidated specs (GPT-5.2)}} \\
\cline{3-11}
 &
& \textbf{code-and-specification} & \textbf{specification-only} & \textbf{code-only}
& \textbf{code-and-specification} & \textbf{specification-only} & \textbf{code-only}
& \textbf{code-and-specification} & \textbf{specification-only} & \textbf{code-only} \\
\hline
MI (SEI) & $\uparrow$
& 44.27 & \cellcolor{gray!25}\textbf{48.57} & 38.85
& \cellcolor{gray!25}\textbf{46.03} & 44.63 & 37.86
& \cellcolor{gray!25}\textbf{45.52} & 45.39 & 38.18 \\
Cyclomatic (sum) & $\downarrow$
& \cellcolor{gray!25}\textbf{14.11} & 15.04 & 16.11
& \cellcolor{gray!25}\textbf{13.44} & 13.86 & 16.46
& \cellcolor{gray!25}\textbf{13.80} & 14.03 & 16.53 \\
Cognitive (sum) & $\downarrow$
& 18.20 & \cellcolor{gray!25}\textbf{17.19} & 20.77
& 17.11 & \cellcolor{gray!25}\textbf{16.92} & 20.45
& 17.44 & \cellcolor{gray!25}\textbf{17.12} & 21.16 \\
Halstead Volume & $\downarrow$
& 2023.19 & \cellcolor{gray!25}\textbf{1831.32} & 2411.87
& 1971.32 & \cellcolor{gray!25}\textbf{1915.54} & 2451.65
& 2017.09 & \cellcolor{gray!25}\textbf{1912.99} & 2449.90 \\
Halstead Effort & $\downarrow$
& 114766.35 & \cellcolor{gray!25}\textbf{89030.40} & 160816.71
& 108583.09 & \cellcolor{gray!25}\textbf{88795.41} & 166571.72
& 112584.04 & \cellcolor{gray!25}\textbf{89475.35} & 165057.89 \\
SLOC & $\downarrow$
& 70.25 & \cellcolor{gray!25}\textbf{69.50} & 81.70
& \cellcolor{gray!25}\textbf{67.88} & 70.01 & 82.59
& 70.29 & \cellcolor{gray!25}\textbf{70.25} & 83.86 \\
PLOC & $\downarrow$
& 53.56 & \cellcolor{gray!25}\textbf{49.37} & 63.19
& 51.33 & \cellcolor{gray!25}\textbf{51.00} & 64.12
& 52.97 & \cellcolor{gray!25}\textbf{51.14} & 65.07 \\
LLOC & $\downarrow$
& 38.59 & \cellcolor{gray!25}\textbf{32.90} & 46.90
& 37.10 & \cellcolor{gray!25}\textbf{35.48} & 47.83
& 38.08 & \cellcolor{gray!25}\textbf{34.95} & 47.45 \\
Comments (CLOC) & $\uparrow$
& 7.39 & \cellcolor{gray!25}\textbf{11.39} & 7.92
& 7.71 & \cellcolor{gray!25}\textbf{9.38} & 7.67
& 8.29 & \cellcolor{gray!25}\textbf{9.70} & 8.19 \\
Blank Lines & $\downarrow$
& 9.93 & \cellcolor{gray!25}\textbf{9.19} & 11.17
& \cellcolor{gray!25}\textbf{9.63} & 10.36 & 11.31
& \cellcolor{gray!25}\textbf{9.82} & 10.14 & 11.21 \\
\hline
\end{tabular}
}

\caption{Comparison of code quality metrics across GPT-5.2-based translation configurations for test-passing translations only. $\uparrow$ indicates higher is better, and $\downarrow$ indicates lower is better. The best values per configuration are highlighted in bold. Group headers give the translation model, the specification type, and (in parentheses) the model that generated the specifications.}
\label{tab:gpt52_comparison}
\end{table*}

\begin{table}[!t]
\centering
\resizebox{\columnwidth}{!}{
\begin{tabular}{l l c c c}
\hline
\textbf{Metric} & \textbf{Pref.}
& \multicolumn{3}{c}{\textbf{GPT-5.2 + test-validated specs (GPT-5.2)}} \\
\cline{3-5}
 & 
& \textbf{code-and-specification} & \textbf{specification-only} & \textbf{code-only} \\
\hline
MI (SEI) & $\uparrow$
& \cellcolor{gray!25}\textbf{38.96} & 37.89 & 30.94 \\
Cyclomatic (sum) & $\downarrow$
& \cellcolor{gray!25}\textbf{16.59} & 17.16 & 19.66 \\
Cognitive (sum) & $\downarrow$
& \cellcolor{gray!25}\textbf{21.44} & 22.45 & 24.55 \\
Halstead Volume & $\downarrow$
& 2435.78 & \cellcolor{gray!25}\textbf{2410.57} & 2960.94 \\
Halstead Effort & $\downarrow$
& 147126.90 & \cellcolor{gray!25}\textbf{131948.86} & 209261.98 \\
SLOC & $\downarrow$
& \cellcolor{gray!25}\textbf{82.13} & 85.00 & 97.69 \\
PLOC & $\downarrow$
& \cellcolor{gray!25}\textbf{62.75} & 62.81 & 76.89 \\
LLOC & $\downarrow$
& 45.84 & \cellcolor{gray!25}\textbf{44.54} & 57.29 \\
Comments (CLOC) & $\uparrow$
& 9.13 & \cellcolor{gray!25}\textbf{11.02} & 8.68 \\
Blank Lines & $\downarrow$
& \cellcolor{gray!25}\textbf{11.38} & 12.17 & 12.82 \\
\hline
\end{tabular}
}
\caption{Comparison of code quality metrics for Translation with Validated Specs using GPT-5.2 in test-passing translations only. $\uparrow$ indicates higher is better, and $\downarrow$ indicates lower is better. The best values are highlighted in bold.}
\label{tab:gpt52_validated_specs}
\end{table}

\paragraph{\textbf{Analysis of Code Quality Metrics for Test-Validated Specifications}}
Table~\ref{tab:gpt52_validated_specs} presents code quality metrics for the  test-validated specification experiment using GPT-5.2, evaluated across 285 
test-passing translations drawn from Table~\ref{fig:gpt_5_2_using_gpt_5_2_validated_specs_status}.
Unlike the results in Table~\ref{tab:gpt52_comparison}, no single modality 
dominates across all metrics: \textit{code-and-specification} leads on structural and complexity measures, while \textit{specification-only} produces more compact and better-documented output. Both, however, consistently outperform \textit{code-only}, which yields the highest complexity values, the largest implementations, and the lowest maintainability score (30.94) across the board.

The \textit{code-and-specification} configuration achieves the highest MI~(SEI) score (38.96), the lowest cyclomatic complexity (16.59) and cognitive complexity (21.44), and the smallest code size (SLOC 82.13, PLOC 62.75). This pattern is consistent with our qualitative analysis (Section~\ref{sec:translation-differences}): access to the original source constrains the model from introducing additional abstractions, such as defensive bounds checks, extra helper functions, and idiomatic wrappers, which can inflate both size and branching complexity. For practitioners prioritizing structural simplicity and control-flow preservation, this Modality is the stronger choice.

The \textit{specification-only} configuration produces the lowest Halstead Volume (2410.57), Halstead Effort (131{,}948.86), and logical line count (44.54), alongside the highest comment density (11.02), yielding less code, better documented, at the cost of marginally higher structural complexity. 
This trade-off is consistent across transformation categories, with the additional abstractions introduced by specification-only, such as defensive bounds checks and Option wrapping, explaining the slight complexity overhead. For practitioners prioritizing readability and compactness, this modality is 
preferable.

The choice between them, however, has modest practical consequences. The \textit{code-only} baseline makes the decision easy by contrast: it produces the highest complexity values, the largest implementations, and an MI~(SEI) of just 30.94, well below both specification-based configurations and far into the low-maintainability range~\cite{herivcko2023exploring}, reinforcing that some form of specification is a prerequisite for structurally sound translation.

To confirm statistical significance, we applied the Friedman test~\cite{cad058f0-09f2-35e2-b8ad-9038985e0961} with post-hoc Wilcoxon signed-rank~\cite{c4091bd3-d888-3152-8886-c284bf66a93a} tests and Holm-Bonferroni correction~\cite{29def780-e117-38f0-8afb-edf384af3fad}. All metrics have significant differences, except Closures ($\chi^2(2) = 1.58,\ p = 0.454$).

For size metrics, code-only produced substantially more code than both specification-based settings across PLOC ($\chi^2(2) = 98.64$), LLOC ($\chi^2(2) = 103.70$), and SLOC ($\chi^2(2) = 88.34$) (all $p < 0.05$, large effect sizes $r > 0.56$), while the combined and specification-only settings were largely comparable. Similarly, for complexity metrics, code-only yielded the highest Cyclomatic Complexity ($\chi^2(2) = 66.00$), Cognitive Complexity ($\chi^2(2) = 35.25$), Exits ($\chi^2(2) = 48.80$), and Functions ($\chi^2(2) = 27.46$) (all $p < 0.05$), with the combined approach consistently producing the lowest values. Halstead Effort ($\chi^2(2) = 117.77$) and Halstead Volume ($\chi^2(2) = 153.23$) followed the same pattern, with code-only significantly exceeding both specification-based settings (all $p < 0.05$, $r > 0.52$). For Blank Lines ($\chi^2(2) = 45.50$), the combined approach produced fewer than both alternatives, while specification-only and code-only did not differ significantly. The Comments metric ($\chi^2(2) = 13.92$) was the only metric in which specification-only led, producing more comments than both other settings ($p < 0.05$).
For the MI (SEI) ($\chi^2(2) = 98.75$) and MI (original) ($\chi^2(2) = 106.68$), both specification-based settings substantially outperformed code-only (large effect sizes), while the combined approach yielded modest but significantly higher maintainability than specification-only ($p < 0.05$, $r \leq 0.307$).

Across all metrics, code-only consistently and substantially underperformed both specification-based approaches. The combined and specification-only settings were largely comparable, confirming that specifications are the primary driver of translation quality, with their combination with code offering only marginal additional gains in maintainability.


\begin{tcolorbox}[
  title=\textbf{RQ2 Answer},
  fonttitle=\bfseries,
  colback=white,
  boxrule=0.5pt,
  arc=0.3em,
  boxsep=1mm,
  left=1mm,
  right=1mm,
  top=1mm,
  bottom=1mm,
]
Code-only translation achieves the highest correctness (98.6\%) but the lowest code quality. Specification-only produces the most idiomatic and maintainable Rust, with approximately $6\times$ fewer unsafe blocks, while code-and-specification offers the best overall balance, achieving 98.3\% correctness with consistently higher-quality code.
\end{tcolorbox}

\subsection{RQ3: To what extent do specifications encourage implementation choices that are less constrained by the original source implementation?}

\label{sec:translation-differences}
To answer this question, we conducted a detailed manual analysis of the 285 programs that passed testing across all three settings. We examined each program by comparing the original C code with its corresponding translations, noting structural and semantic differences between the code-only and specification-based outputs. We identified recurring patterns and counted how frequently each appeared across all programs to assess whether the observations were systematic rather than incidental. This analysis reveals two consistent and complementary patterns, described in what follows. 

\paragraph{\textbf{What code-only translation preserves from C}}

When the LLM has direct access to the source code, it tends to mirror C's implementation choices rather than producing implementations free of C's structural choices. This manifests in several recurring ways. Variable names from the original C source, including single-letter loop counters and opaque intermediate names, are carried over into the Rust output in 98\% of code-only translations, which makes the translated code harder to read. Integer types follow C conventions: \texttt{i32} is used where \texttt{usize} or \texttt{i64} would be more natural in Rust in 86\% of code-only translations, and signed wrapping arithmetic is sometimes added explicitly to replicate C's undefined behavior on overflow \cite{p00013}. Loop structures, including manual index manipulation and while loops with mutable counters, are preserved rather than replaced by Rust's iterator idioms \cite{p00015} in 91\% of code-only translations. Fixed-size arrays matching C's stack-allocated dimensions appear in 70\% of code-only translations in place of more flexible \texttt{Vec} allocations \cite{p00071}. When C uses global mutable state, hand-rolled data structures, or low-level I/O patterns, the code-only translation reproduces these with varying degrees of fidelity, sometimes resorting to \texttt{unsafe} blocks to match C's unguarded memory access patterns \cite{p00244}. In 21.7\% cases, redundant or dead variables are also reproduced in the output rather than corrected \cite{p00023}.

\paragraph{\textbf{What specification-based translation abstracts and improves}}

When the LLM works from a specification rather than source code, it is no longer anchored to C's implementation decisions and instead produces implementations that reflect the problem's semantic rather than C's structure. This leads to a consistent set of improvements. Algorithms are often replaced by cleaner equivalents: a hand-rolled Newton's method square root loop becomes a single \texttt{.sqrt()} call \cite{p00010}, a manual bubble sort becomes \texttt{sort\_by} \cite{p00124}, and similar substitutions occur in 90.7\% of specification-only translations and 67.6\% of code-and-specification translations, compared to just 1.7\% of code-only translations. Data structures are chosen to match the problem's semantics rather than C's memory layout: parallel raw arrays become named structs with meaningful field names \cite{p00453}, fixed-size hash tables with manual collision handling become \texttt{HashSet} \cite{p00949}, and manual heap implementations are replaced by \texttt{BinaryHeap} from the standard library \cite{p00244}, with such upgrades appearing in 78.7\% of specification-only and 66.2\% of code-and-specification translations versus 4\% of code-only. Type choices improve: counts described as non-negative in the specification are represented as \texttt{usize} or \texttt{u64} rather than \texttt{i32} in 98.3\% of specification-only translations, and integer sentinel values like -1 become \texttt{Option} types~\cite{p00261} in 58.7\% of specification-only and 45.5\% of code-and-specification translations, compared to 1.3\% of code-only. Unsafe code largely disappears: when the specification states that inputs satisfy certain bounds as preconditions, the LLM no longer needs to emulate C's unguarded indexing and can use safe Rust indexing directly~\cite{p00013}. As shown in Table~\ref{tab:unsafe}, specification-only translation reduces unsafe blocks by roughly 6 times compared to code-only, with code-and-specification in between. Unlike prior work such as SpecTra ~\cite{nitin2024spectra}, which evaluates specifications alongside source code and measures correctness alone, these results show that specifications used in isolation can meaningfully shift translation away from C's structural choices.

\begin{tcolorbox}[
  title=\textbf{RQ3 Answer},
  fonttitle=\bfseries,
  colback=white,
  boxrule=0.5pt,
  arc=0.3em,
  boxsep=1mm,
  left=1mm,
  right=1mm,
  top=1mm,
  bottom=1mm,
]
Specification-based translation produces implementations that are free from C’s structural choices, replacing C-style constructs with idiomatic Rust equivalents in the majority of programs: algorithm substitution occurs in 90.7\% of specification-only translations versus 1.7\% of code-only, and data structure upgrades appear in 78.7\% versus 4\%. Code-only translation, by contrast, mirrors C's implementation choices across all categories. These results confirm that natural language specifications shift LLM translation away from syntactic mimicry toward implementations that are less constrained by the original source.

\end{tcolorbox}

\subsection{RQ4: What are the cost–quality trade-offs of specification-based translation pipelines?}

\begin{figure}[!tp]
    \centering
    \includegraphics[width=\linewidth,height=0.99\textheight,keepaspectratio]{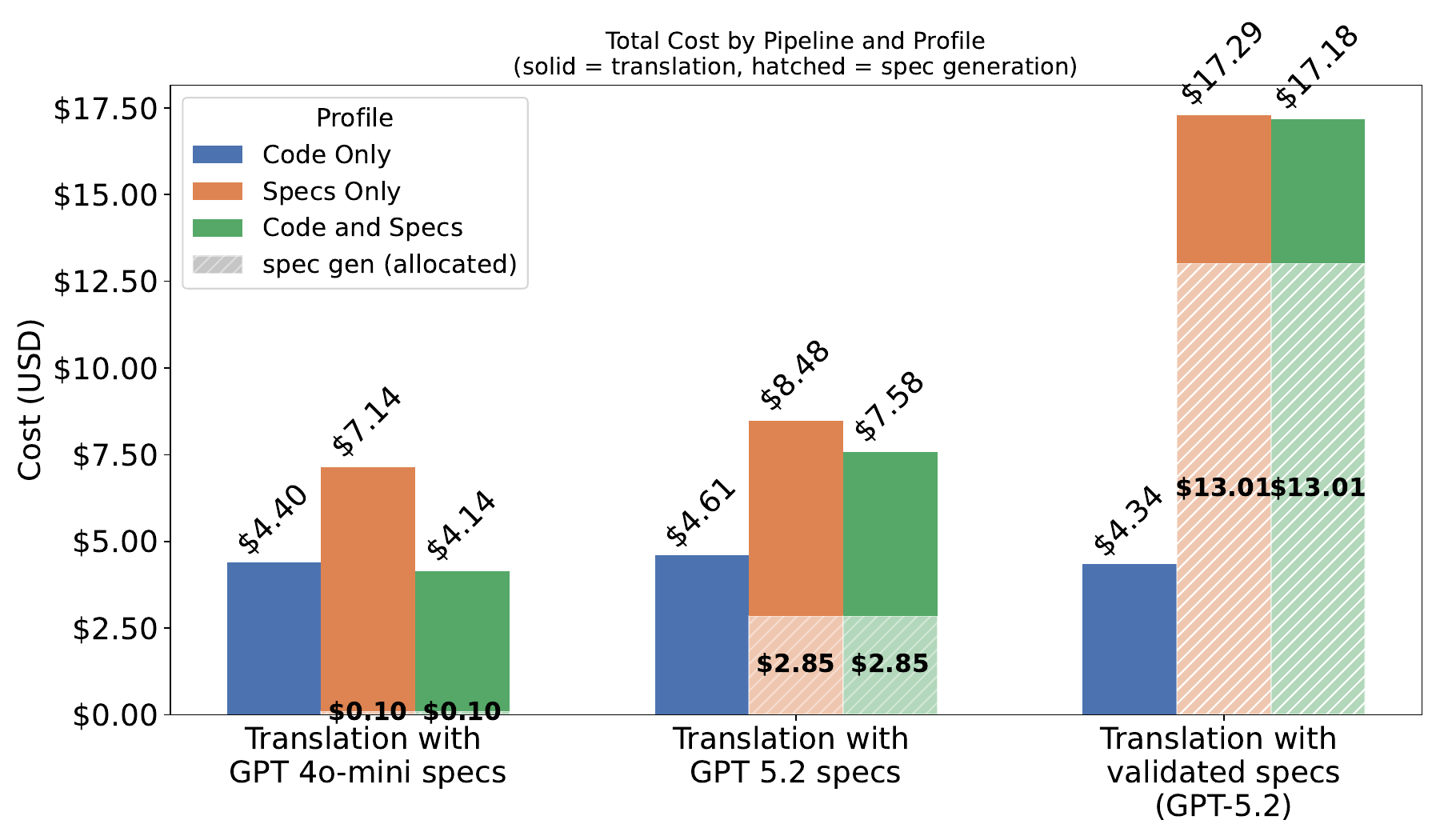}
    \caption{Cost analysis for different translation profiles}
    \label{fig:cost_analysis}
\end{figure}

Figure~\ref{fig:cost_analysis} presents the translation cost using Tiktoken~\cite{tiktoken}. The \textit{code-only} profile remains stable 
across pipelines (\$4.34--\$4.61), while \textit{specification-only} 
consistently incurs the highest cost due to more iterations required to 
produce test-passing translations. The test-validated specification pipeline 
substantially increases overall cost, spending \$13.01 on validation alone 
compared to \$2.85 for unvalidated specs, bringing total translation costs 
to \$17.29 and \$17.18 for specification-only and code-and-specification 
respectively.

Interpreted alongside our analysis of previous RQs, these cost differentials reveal important trade-offs. Despite its low cost (\$4.45 average), \textit{code-only} consistently yields the lowest maintainability and highest complexity, negating its cost advantage. The test-validated specification pipeline, despite its high cost, does not yield commensurate improvements in structural quality compared with unvalidated specifications, making GPT-5.2-generated specifications the most cost-efficient operating point.
Critically, structural metrics alone do not capture translation fidelity. As discussed in Section~\ref{sec:quality_metrics}, structurally superior translations are not necessarily functionally correct: GPT-4o-mini-based specifications produce compact, maintainable code but exhibit elevated test failure rates due to under-specification. The cost of validation should be understood as an investment in functional correctness rather than structural quality. Thus, practitioners optimizing purely for cost risk produce structurally sound but semantically deficient translations. This is important when applying specification-based translation in the context of security-critical software migration.

\begin{tcolorbox}[
  title=\textbf{RQ4 Answer},
  fonttitle=\bfseries,
  colback=white,
  boxrule=0.5pt,
  arc=0.3em,
  boxsep=1mm,
  left=1mm,
  right=1mm,
  top=1mm,
  bottom=1mm,
]

The right pipeline depends on what the migration prioritizes:
\begin{itemize}
  \item \textbf{Correctness and cost only:} code-only (98.7\% pass rate, \$4.34), if code quality is not a concern; it produces the most unsafe blocks and Clippy warnings.
  \item \textbf{Correctness and code quality:} code-and-specification with test-validated specifications (98.3\%, \$17.18), which keeps near code-only correctness with fewer unsafe blocks and warnings, at about 4$\times$ the cost.
  \item \textbf{Code quality first:} specification-only with test-validated specifications (95.3\%, \$17.29), which produces the fewest unsafe blocks and Clippy warnings.
  \item \textbf{Limited budget:} code-and-specification with unvalidated GPT-5.2 specifications (96.0\%, \$7.58), which gives most of the quality gains at less than half the cost of the test-validated pipeline. Specification-only should not be used without validation (86.3\%).
\end{itemize}
\end{tcolorbox}

\section{Threats to Validity}
\label{sec:threats}

This study has some limitations that should be considered when interpreting the results. We discuss the primary threats to internal, construct, external, and conclusion validity, together with the measures taken to mitigate their impact.

\textbf{Internal Validity.}
Specification generation, validation, and translation are executed as independent, stateless API calls with no shared context, preventing information leakage across stages. In the specification-only setting, the model receives only the generated specification, ensuring a clean evaluation of specification sufficiency. Specification correctness is assessed through back-translation and execution against ground-truth test cases rather than LLM self-assessment. While back-translation against test cases does not guarantee full semantic completeness, this test-based validation approach is consistent with how correctness is assessed in prior work~\cite{nitin2024spectra}. Validation and evaluation use the same test cases. Although the refinement prompt receives only a diagnosis rather than the test cases themselves, the diagnosis is derived from failing tests, and a specification is accepted only once it passes them. Test-validated specifications may therefore be tuned to the tested behaviors, which could inflate pass rates in the specification-based settings compared with a held-out test set. The Rust repair loop uses execution feedback from the same tests in all three settings, so this concern applies only to the additional specification refinement step. LLM non-determinism and the $k{=}3$ repair loop may introduce variance across runs, which we partially mitigate by evaluating two models with different capabilities.

\textbf{Construct Validity.}
Structural quality metrics are computed using \texttt{rust-code-analysis}, which analyzes the Rust AST but does not capture Rust-specific semantics such as ownership, lifetimes, or trait dispatch; consequently, maintainability and complexity should be interpreted as structural approximations. Likewise, Clippy warnings serve only as a proxy for idiomaticity, which we complement with unsafe-construct metrics. Translation cost is estimated using Tiktoken rather than actual API billing, but the methodology is applied consistently across all settings, preserving the validity of relative comparisons.

\textbf{External Validity.}
Our evaluation uses the 300-problem IBM CodeNet subset, whose programs are shorter and more self-contained than industrial C systems. Although this limits generalizability to repository-scale migration, it enables controlled comparisons and direct comparison with prior work~\cite{nitin2024spectra}. Furthermore, our study is limited to two OpenAI models and a single language pair (C-to-Rust).

\textbf{Conclusion Validity.}
To ensure fair comparisons, quality metrics are computed only on the balanced intersection of programs successfully translated across all three settings (193 for unvalidated specifications and 285 for test-validated specifications), thereby preventing differences in problem coverage from confounding the observed quality metrics.

\section{CONCLUSION}
\label{sec:conclusion}
This paper investigated whether natural language specifications can improve the quality of LLM-based C-to-Rust translation beyond functional correctness. We compared three translation settings (code-only, specification-only, and code-and-specification) across 300 C programs, evaluating correctness, maintainability, complexity, idiomaticity, unsafe-construct usage, and cost. Our results show a clear trade-off. Code-only translation achieves the highest correctness (98.6\%) but consistently produces less maintainable, more complex, and less idiomatic Rust by mirroring C's implementation choices. In contrast, specification-based translation frees the LLM from C's structural choices, producing safer and more idiomatic Rust, reducing unsafe blocks by approximately 6$\times$ and Clippy style warnings by 2.4$\times$. While specification-only translation initially lowers correctness to 86\%, our iterative test-based validation and refinement process increases it to 95.3\%, demonstrating that specification quality is critical to translation effectiveness. Code-and-specification provides the best overall balance, combining competitive correctness (98.3\%) with substantial improvements in code quality.

Overall, our findings demonstrate that test-validated natural language specifications are an effective intermediate representation for software modernization, enabling LLMs to generate Rust that is not only functionally correct but also safer, more maintainable, and more idiomatic. Future work includes extending this approach to repository-level systems and additional source-target language pairs.

\section*{Data Availability}
To enable further research and replication, the data and code used in our study are publicly available~\cite{supplementary}.

\bibliographystyle{IEEEtran}
\bibliography{main}
\end{document}